\documentstyle[prd,aps,psfig,floats,epsfig]{revtex}

\begin{document} 
\bibliographystyle{prsty} 
%\bibliographystyle{apsrev} 
%\bibliographystyle{h-physrev} 
%\draft
\input epsf
\twocolumn[\hsize\textwidth\columnwidth\hsize\csname 
@twocolumnfalse\endcsname 
\title{Principles, Progress and Problems in Inflationary Cosmology}
\author{Robert H. Brandenberger}
\address{Physics Department, Brown University, Providence, RI, 02912, USA \\ 
and \\
Theory Division, CERN, CH-1211 Gen\`eve 23, Switzerland}
\date{\today} 
\maketitle 
 
\begin{abstract} 
Inflationary cosmology has become one of the cornerstones of 
modern cosmology. Inflation was the first theory within which it was 
possible to make predictions about the structure of the Universe on large 
scales, based on causal physics. The development of the inflationary Universe 
scenario has opened up a new and extremely promising avenue for connecting 
fundamental physics with experiment. This article summarizes the 
principles of inflationary cosmology, discusses progress in the field,
focusing in particular on the mechanism by
which initial quantum vacuum fluctuations develop into the seeds for
the large-scale structure in the Universe, and highlights the important
unsolved problems of the scenario. The case is made that new input
from fundamental physics is needed in order to solve these problems,
and that thus early Universe cosmology can become the testing ground
for trans-Planckian physics.
\end{abstract}

\vspace{0.35cm} 
] 

\section{Introduction}

With the recent high-accuracy measurements of the spectrum of the
cosmic microwave background (CMB) (see Fig. 1 \cite{Scott}), 
cosmology has become
a quantitative science. There is now a wealth of new data on the
structure of the Universe as deduced from precision maps of the
cosmic microwave background anisotropies, from cosmological redshift
surveys, from redshift-magnitude diagrams of supernovae, and from
many other sources. Standard Big Bang (SBB) cosmology provides
the framework for describing the present data. The interpretation
and explanation of the existing data, however, requires us to go
beyond SBB cosmology and to consider scenarios like the Inflationary
Universe in which space-time evolution in the very early Universe
differs in crucial ways from what is predicted by the SBB theory. Since
inflationary cosmology at later times smoothly connects with the SBB
picture, we must begin this article with a short review of the
framework of SBB cosmology.

Standard big bang cosmology rests on three theoretical pillars: the
cosmological principle, Einstein's general theory of relativity, 
and the assumption that matter is a classical perfect fluid.
The cosmological principle concerns the symmetry of space-time
and states that on large distance scales space
is homogeneous and isotropic. This implies that the metric of space-time can 
be written in Friedmann-Robertson-Walker (FRW) form. For simplicity, we
consider the case of a spatially flat Universe:
\begin{equation}
ds^2 \, = \, dt^2 - a(t)^2 
\left[dr^2 + r^2 (d \vartheta^2 + \sin^2 \vartheta d\varphi^2) \right] \, . 
\end{equation}
Here, $t, r, \theta$ and $\varphi$ are the space-time coordinates, and
$ds$ gives the proper time between events in space-time.
The coordinates $r, \vartheta$ and $\varphi$ are ``comoving'' spherical 
coordinates, and $t$ is the physical time coordinate. 
Space-time curves with constant comoving coordinates 
correspond to the trajectories of particles at rest. If the Universe is 
expanding, i.e. the scale factor $a(t)$ is increasing, then the physical 
distance $\Delta x_p(t)$ between two points at rest with fixed 
comoving distance $\Delta x_c$ grows as
$\Delta x_p = a(t) \Delta x_c$.

\begin{figure}[htb]
\begin{center}
%\leavevmode
%\epsfysize=10.5cm \epsfbox{ispectrum_01.eps}
\includegraphics[width=7.5cm,angle=+90]{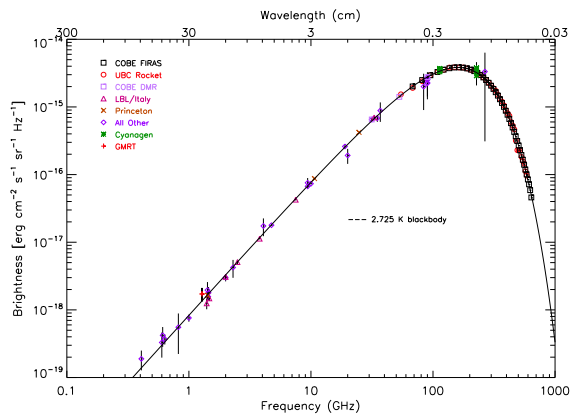}
\caption{Compilation of the spectrum of the CMB. 
In the region around the peak (the region probed with greatest
precision by the COBE satellite) the error bars are smaller than
the size of the data points.}
\end{center}
\end{figure}
  
The dynamics of an expanding Universe  is determined by the Einstein equations,
which relate the expansion  rate to the matter content, specifically to the
energy density $\rho$ and pressure $p$.  For a Universe 
obeying the cosmological principle (and neglecting the possible presence
of a cosmological constant) the Einstein equations reduce to the 
Friedmann-Robertson-Walker (FRW) equations
\begin{eqnarray} \label{FRW}
\left( {\dot a \over a} \right)^2 - {k\over a^2} \, 
&=& \, {8 \pi G\over 3 } \rho \\
\dot \rho \, &=& \, - 3 H (\rho + p) \, ,
\end{eqnarray}
where $H = {\dot a}/a$ is the Hubble expansion rate,
and an overdot denotes the derivative with respect to time $t$.

The third key assumption of standard cosmology is that matter is described by
a classical ideal gas with some equation of state which is conveniently
parametrized in the form
$p = w \rho$ which some constant $w$. 
For cold matter, pressure is negligible and hence $w = 0$.  
For radiation we have $w = {1/3}$. In standard cosmology, the Universe
is a mixture of cold matter and radiation, the former dominating at
late times, the latter dominating in the early Universe. In this
case, the FRW equations can be solved exactly, with the result that
the Universe had to be born in a ``Big Bang'' singularity.

SBB cosmology explains Hubble's redshift-distance relationship, and it
explains the abundances of the light elements Helium, Deuterium
and Lithium. These nucleosynthesis predictions of the SBB model depend 
on a single free parameter (which is the present
ratio of the number densities of baryons to photons), and thus
the fact that the abundances of more than one element can be matched
by adjusting this ratio is remarkable. Most importantly,
SBB cosmology predicts the existence and black-body spectral
distribution of a microwave cosmic background radiation, the CMB. The
precision measurement \cite{Gush,COBE} of the spectrum of the CMB is a 
triumph for SBB cosmology (see Fig. 1).

\begin{figure} %\label{horprob}
\begin{center}
\leavevmode
\epsfysize=4.5cm \epsfbox{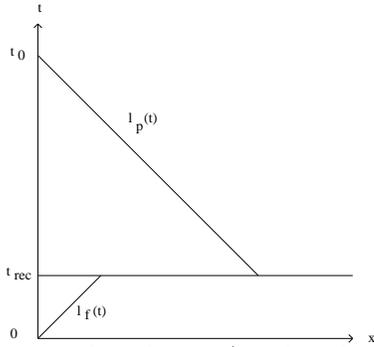}
\caption{
A space-time diagram (physical distance $x_p$ versus time $t$)
illustrating the homogeneity problem: the past light cone $\ell_p (t)$ at the
time $t_{rec}$ of last scattering  is much larger than the forward light cone
$\ell_f (t)$ at $t_{rec}$.}
\end{center}
\end{figure}

However, the triumph of the CMB also leads to one of several major
problems for SBB cosmology: within the context of this theory, there
is no explanation for the high degree of isotropy of the radiation.
As is illustrated in Fig. 2, at the time the microwave radiation last
scattered (which occurred when the temperature was about a factor $10^3$
of the present temperature), the maximal distance which light could
have communicated information starting at the Big Bang (the
forward light cone) is much smaller
than the distance over which the microwave photons are observed to have
the same temperature (the past light cone). 
This is the famous ``horizon problem'' of
SBB cosmology. Within the context of SBB cosmology it is also a
mystery why the Universe today is observed to be approximately spatially
flat, since a spatially flat Universe is an unstable fixed point of the
FRW equations in an expanding phase. This problem is called the ``flatness
problem''. Finally, as illustrated in Fig. 3, within standard cosmology
there is no causal mechanism which can explain the nonrandom distribution
of the seed inhomogeneities which develop into the present-day 
large-scale structure. This constitutes the ``formation of structure
problem''. Under the assumption that only gravity is responsible for
the development of inhomogeneities on cosmological scales, the seeds for
fluctuations
must have been correlated at the time $t_{eq}$, the time when the
energy densities in cold matter and radiation were equal (which
occurred when the Universe was about $10^{-4}$ of its present size) 
when inhomogeneities
can first start to grow by gravitational instability. But, at that time,
the forward light cone was smaller than the separation between the seeds.

\begin{figure}
\begin{center}
\leavevmode
\epsfysize=5.0cm \epsfbox{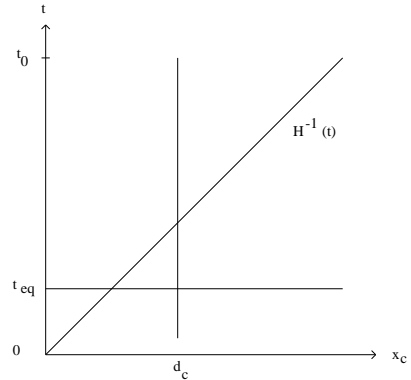}
\caption{A sketch (conformal separation vs. time) of the formation of
structure problem: the comoving separation $d_c$ between two clusters is larger
than the forward light cone at time $t_{eq}$.}
\end{center}
\end{figure}

Standard Big Bang cosmology is also internally inconsistent as
a theory of the very early Universe. We know that at very high
energy densities which the theory predicts in the initial stages,
the description of matter as a classical ideal gas is invalid. Since
quantum field theory is a better description of matter at high
energies,
this naturally leads us to consider quantum field
theory as the description of matter which must take over in the
very early Universe, and this realization led to the
discovery of the inflationary scenario. 
What follows is a brief overview of principles,
progress and problems in inflationary cosmology. For more details,
the reader is referred to \cite{RHB}.

\section{The Inflationary Universe Scenario}

The inflationary Universe scenario \cite{Guth} is based on the simple
hypothesis that there was a time interval 
in the early Universe beginning at some time $t_i$ and ending at 
a later time $t_R$ (the ``reheating time") during
which the scale factor is exponentially expanding. Such a period is 
called  ``de Sitter" or ``inflationary".  The success of 
Big Bang nucleosynthesis demands that this time interval was long
before the time of nucleosynthesis. It turns out that during the
inflationary phase, the energy of matter is stored in some new form
(see below). At the time $t_R$ of reheating, 
all this energy is released as thermal energy.  This is a
nonadiabatic process during which the entropy of the Universe 
increases by a large factor.   

Independent of any specific realization, the above simple hypothesis
of inflation leads immediately to possible solutions of the horizon, flatness
and formation of structure problems. Fig. 4 is a sketch of how a period of 
inflation can solve the horizon
problem. During
inflation, the forward light cone increases exponentially compared to a model
without inflation, whereas the region over which isotropy
is observed is not affected. Hence, provided inflation lasts 
sufficiently long, the forward light cone at the time of last
scattering of CMB photons can be made larger than the region from which
the microwave photons are reaching us today.  

\begin{figure}
\begin{center}
\leavevmode
\epsfysize=6cm \epsfbox{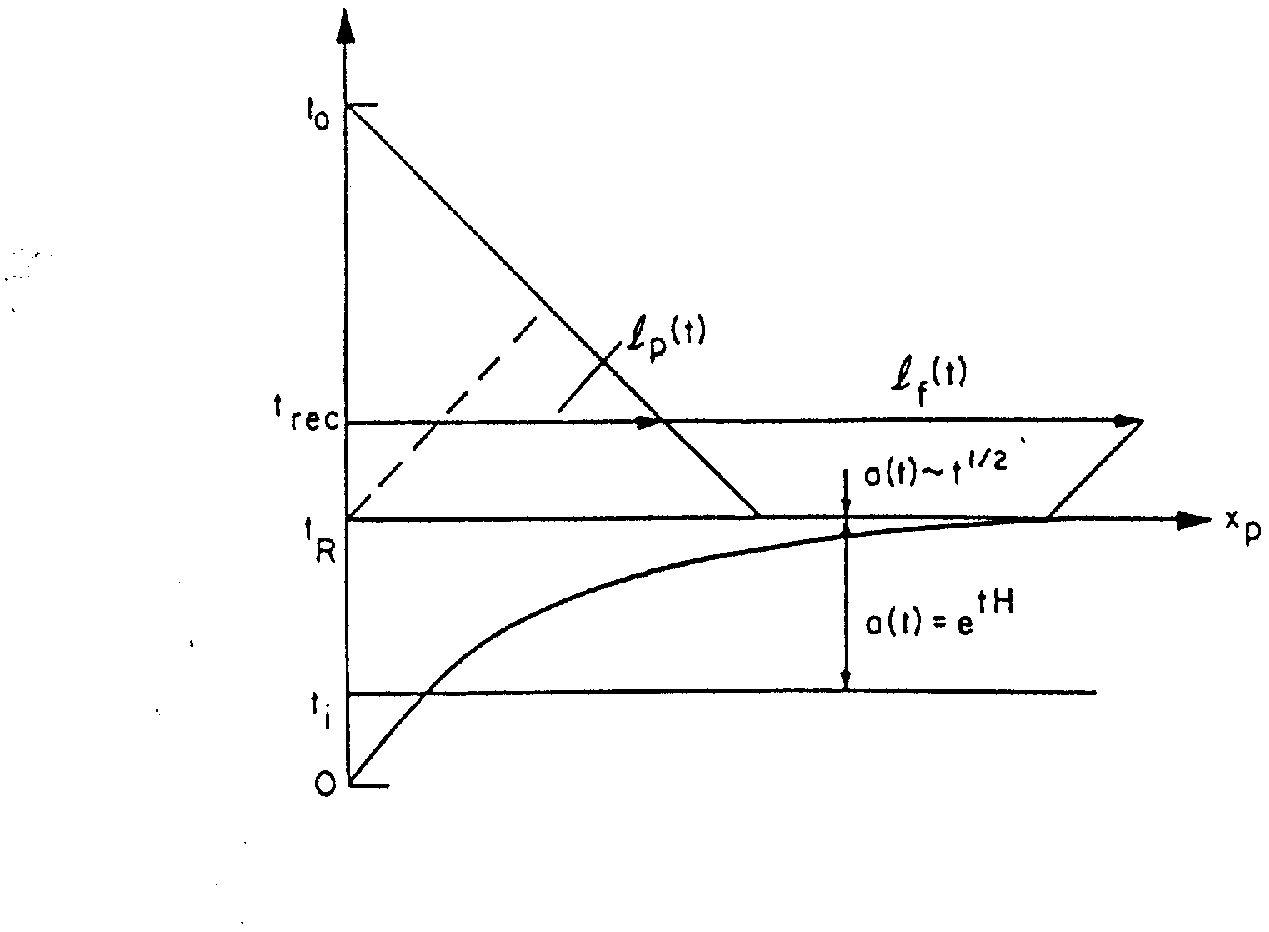}
\caption{
Sketch (physical distance $x_p$ vs.time $t$) of the
solution of the homogeneity problem. During inflation, the forward light cone
$l_f(t)$ is expanded exponentially when measured in physical coordinates.
Hence, it does not require many e-foldings of inflation in order that $l_f(t)$
becomes larger than the past light cone $l_p(t)$ at the time $t_{\rm rec}$ 
of last scattering. The
dashed line is the forward light cone without inflation.}
\end{center}
\end{figure}  

Inflation also can solve the flatness problem. The key point is
that at the time of reheating, the entropy of the Universe increases
by a large factor, and this drives the Universe towards
spatial flatness, as can be seen from the FRW equations.  In fact, one
of the key predictions of inflationary cosmology is that the Universe
should be spatially flat to great accuracy (although it is possible
to construct special models of inflation which produce any given
deviation from spatial flatness).

Most importantly, inflation provides a mechanism which in a causal way
generates the primordial perturbations required to explain the
nonrandom distribution of matter on the scales of galaxies and galaxy clusters,
and to produce small-amplitude anisotropies in the CMB. At any given
time, microphysics can act coherently on scales up to the Hubble radius 
$H^{-1}(t)$. The key point is that during the inflationary period the Hubble
radius is constant, whereas the physical length scale associated with
fluctuations which have fixed comoving scale increases exponentially.
Thus, as is depicted in Fig. 5, provided that the inflationary period
is sufficiently long, all comoving scales of cosmological interest
today had a physical wavelength smaller than the Hubble radius in the
early stages of inflation. Thus, it is possible without violating
causality to have a mechanism which generates microscopic-scale
fluctuations during the period of inflation whose wavelengths then
get stretched exponentially so that they can become the seeds for
structure in the present Universe.

\begin{figure}
\begin{center}
\leavevmode
\epsfysize=6cm \epsfbox{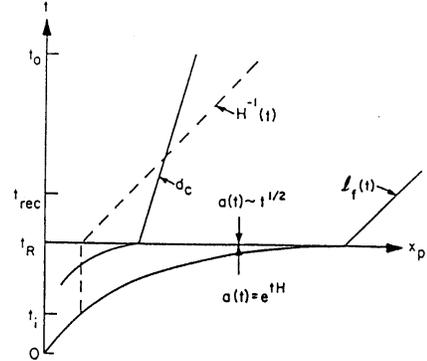}
\caption{
A sketch (physical coordinates vs. time) of the
solution of the formation of structure problem. Provided that the period of
inflation is sufficiently long, the separation $d_c$ between two galaxy
clusters is at all times smaller than the forward light cone. The dashed line
indicates the Hubble radius. Note that $d_c$ starts out smaller than the Hubble
radius $H^{-1}(t)$, crosses it during the de Sitter period, and then 
reenters it at late times.}
\end{center}
\end{figure}

As will be discussed in Section 4, the density perturbations produced during
inflation are due to quantum fluctuations in the matter and gravitational
fields$^{\cite{Mukh80,Lukash}}$. These fluctuations are
continously generated during the period of inflation. Once the physical
wavelength equals the Hubble radius, the vacuum oscillations
freeze out, the quantum state is squeezed, and the highly squeezed
state at late time appears as a state with a large number of 
particles, representing the density fluctuations at late times.  
The amplitude of these
fluctuations is given by the Hubble expansion rate $H$. Since $H$ is
approximately constant during the period of inflation, this mechanism
of quantum vacuum fluctuations freezing out and becoming classical
density fluctuations leads to the prediction that the spectrum of
fluctuations should be scale invariant, i.e. the physical measure
of the amplitude of fluctuations at late times should be independent
of the wavenumber $k$:
\begin{equation}
k^3 |\delta_k|^2 \, = \, {\rm const} \, ,
\end{equation}
where $\delta_k$ is the fractional energy density fluctuation in
momentum space. Thus, independent of the specific mechanism which
drives inflation, as long as the expansion is almost exponential,
the spectrum of fluctuations is predicted to be almost scale-invariant.
As will be discussed in Section 4, this quantitative prediction of
inflationary cosmology is confirmed by observations. However, as
will be mentioned in Section 5, this prediction may depend on
unstated assumptions about the trans-Planckian physics.

\section{How to Obtain Inflation}

Because in the year 1980, when inflationary cosmology was being
developed, quantum field theory was the best
available description of matter at high energies such as must have
occurred close to the Big Bang, it in retrospect seems obvious to
turn to it for a possible implementation of the inflationary Universe
scenario. Since scalar matter field are supposed to play an
important role in high energy physics, in particular for implementing
the spontaneous breaking of internal gauge symmetries, it is
necessary to consider the role of such fields in cosmology.

If we assume that Einstein's equations remain valid in the early
Universe, then it follows from the FRW equations (\ref{FRW}) that
an equation of state with negative pressure $p \simeq - \rho$ is 
required to obtain exponential inflation. In order to obtain an
accelerated expansion (``generalized inflation''), an
equation of state with $p < - (1/3) \rho$ is needed. In the context
of renormalizable quantum field Lagrangians, it is only scalar
fields which provide the possibility to obtain inflation. From the
action of a scalar quantum field $\varphi$ (in an expanding space-time) it
immediately follows that the energy density and pressure of such
a field are given by
\begin{eqnarray}
\rho \, &=& \, {1 \over 2}({\dot \varphi})^2 + 
{1 \over {2a^2}}(\nabla \varphi)^2 + V(\varphi) \\
p \, &=&  {1 \over 2}({\dot \varphi})^2 -
{1 \over {6a^2}}(\nabla \varphi)^2 - V(\varphi) \, ,
\end{eqnarray}
where $V(\varphi)$ is the potential energy density.
It thus follows that if the scalar field is homogeneous and static, but the 
potential energy positive, then the equation of state $p = - \rho$ necessary 
for exponential inflation results. This is the idea behind potential-driven 
inflation.

Guth's initial hope was that the same scalar field responsible for
the spontaneous symmetry breaking of a unified gauge symmetry could
be the inflaton, the field responsible for inflation. However, this hope
cannot be realized since the potential of such a scalar field is
too steep in order to provide a period of more than a few Hubble
times $H^{-1}$ during which the field kinetic energy remains negligible.
This follows almost immediately by considering the variational equation
of motion for $\varphi$, which in the absence of spatial gradients becomes
\begin{equation} \label{eom}
\ddot \varphi + 3 H \dot \varphi \, = \, - V^\prime (\varphi)\, ,  
\end{equation}
where a prime denotes the derivative with respect to $\varphi$.

Although at the present time there are many models for inflation, there is
no single convincing one. Chaotic inflation \cite{Linde} is a
prototypical model. It assumes the existence of a new scalar field,
the ``inflaton'' with a smooth potential (see Fig. 6). The inflaton
is so weakly coupled that it does not start out in thermal equilibrium
in the early Universe. Most of the phase space of initial conditions
consists of values of $\varphi$ which are large (compared to the Planck
scale). It follows from the equation of motion (\ref{eom}) that
for such initial conditions the scalar field will roll slowly (in the
sense that $\ddot{\varphi} \ll 3 H \dot{\varphi}$) for a long (compared
to $H^{-1}$) time, thus yielding inflation. Once $\varphi$ drops
much below the Planck scale, the kinetic energy begins to dominate,
the field oscillates around its minimum and releases its energy
to regular matter via interaction terms in the Lagrangian.

\begin{figure}
\begin{center}
\leavevmode
\epsfysize=5.5cm \epsfbox{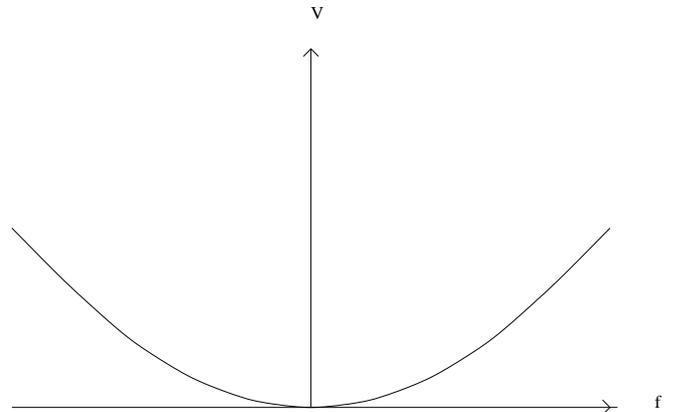}
\caption{Sketch of a  potential $V$ for a scalar field $f$ in
chaotic inflation.}
\end{center}
\end{figure}

\section{Progress in Inflationary Cosmology}

Simple prototypical models of inflation predict a scale-invariant
spectrum of adiabatic fluctuations (adiabatic in this context means
that the energy density fluctuations in all components of matter are
proportional), and this will arise naturally in the context of inflation
if during the phase of inflationary expansion a single scalar field
dominates the dynamics). This prediction of the inflationary
scenario has recently been confirmed to high  accuracy by the
measurements of CMB anisotropies. Microwave anisotropies are
usually quantified by expanding the temperature maps into spherical
harmonics and calculating the power at different values of the
angular quantum number $l$. Fig. 7 \cite{Scott} shows a compilation of
the recent measurements. The predictions of a theory based on
a scale-invariant spectrum of adiabatic fluctuations with a background
cosmology which is taken to be dominated today by a remnant cosmological
constant which contributes $70\%$ to the density required for a spatially
flat Universe (most of the other $30\%$ is believed to be made up of cold
dark matter) fits the data very well, whereas a model with a background
dominated by cold dark matter does not reproduce the observed details
of the spectrum in the region of the oscillations of the spectrum. 

\begin{figure}
\begin{center}
\leavevmode
\epsfysize=8.5cm \epsfbox{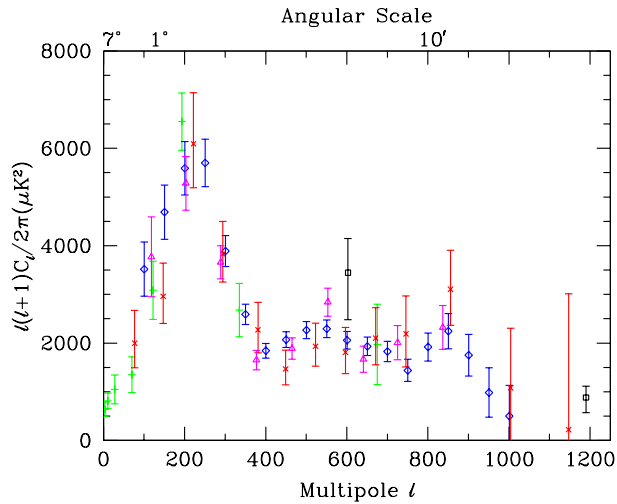}
\caption{Compilation of recent data on the angular power 
spectrum of CMB anisotropies. The green (-) data points
are a compilation of the data prior to March 2000; the others represent
data from the following experiments: Boomerang (2001 release), blue (diamond);
Maxima (2001 release), purple (triangle);
DASI, red (x), CBI, black (square).}
\end{center}
\end{figure}

Note the oscillations in the spectrum at values of $l$ larger than $100$.
These ``acoustic'' oscillations are an imprint of the coherence of the 
primordial fluctuations. Refer to Fig. 8 for a sketch of the mechanism by which
density fluctuations induce CMB anisotropies.
If we imagine the inhomogeneities as a superposition of
plane waves (which evolve independently in the early Universe since their
amplitudes are small), then inflation predicts that these waves start
oscillating with the same phase at the time when the wavelength equals
the Hubble radius. Thus, when measured at the time of last scattering, the
phase of the wave is a periodically varying function of $l$. Maxima and
minima of the waves result in maxima of the temperature anisotropies on
the corresponding angular scales, nodes of the waves result in minima
(see e.g. \cite{Hu} for a detailed discussion of the physics of the
acoustic oscillations).

The theoretical understanding of the theory by which quantum vacuum
fluctuations on microscopic scales evolve to give rise to classical
inhomogeneities on cosmological wavelengths is one of the main areas
of progress in inflationary cosmology. Since it is necessary to
propagate the fluctuations on scales larger than the Hubble radius,
the Newtonian theory of
cosmological perturbations obviously is inapplicable, and a general
relativistic analysis is needed.  On these scales, matter is essentially frozen
in comoving coordinates.  However, space-time fluctuations can still increase
in amplitude. In principle, it is straightforward to work out the general 
relativistic theory of linear fluctuations. One linearizes the Einstein  
and scalar matter field equations about an expanding FRW background cosmology.
Tedious but straightforward algebra gives a set of coupled linear differential
equations for the metric and matter perturbations.

At the level of linear fluctuations, perturbations with different wavelengths
decouple. Hence, the analysis becomes simple if we work in momentum
space. Furthermore, there are several independent degrees
of freedom.
First, there are gravitational waves, space-time perturbations which do
not couple to matter. Next, there are vector perturbations which
correspond to rotational degrees of freedom. Finally, there are
the scalar metric fluctuations, space-time inhomogeneities produced
by matter perturbations. The scalar metric fluctuations are the most
difficult to analyze, in particular since one must be careful to isolate
the physical degrees of freedom from gauge artefacts, modes which correspond
to space-time coordinate transformations. For matter described by a
single scalar field, as is the case in many prototypical inflationary
models, there is only one matter fluctuating degree of freedom. According
to the Einstein equations, matter fluctuations are coupled to
corresponding scalar metric perturbations. There can be no scalar metric
fluctuations without matter inhomogeneities. Thus, restricting attention
to scalar metric fluctuations of one particular wavelength, there is
only one physical degree of freedom. This implies that the analysis
of scalar metric fluctuations is  reducible to the theory of a single 
free scalar field (free because we are dealing with linear fluctuations)
in a time-dependent background (the time-dependence is set by the
background cosmology).
  
\begin{figure}
\begin{center}
\leavevmode
\epsfysize=4.5cm \epsfbox{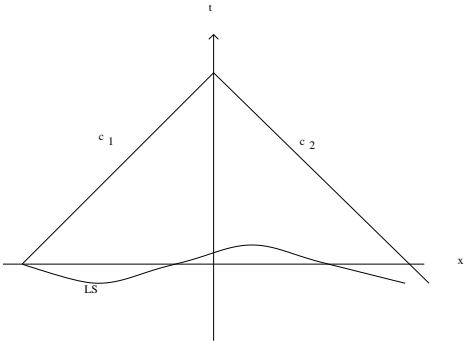}
\caption{Sketch of the geometry which underlies the imprint of
density fluctuations as anisotropies in the cosmic microwave
background. The surface of last scattering (solid line LS) is a surface
of constant temperature. Thus, photons $c_1$ and $c_2$ from different 
directions in
the sky are redshifted by different amounts before reaching us (we
are at the point along the t-axis where the lines labelled $c_1$ and $c_2$ 
hit), and therefore arrive with different temperatures.}
\end{center}
\end{figure}

As mentioned earlier, the primordial fluctuations in an inflationary 
cosmology result from quantum fluctuations.  
At the linearized level, the equations describing 
both gravitational and matter perturbations can be quantized in a consistent 
way (for a detailed review see \cite{MFB}).
The first step of this analysis is to 
expand the gravitational and matter action  to 
quadratic order in the metric and matter fluctuation variables.
Focusing on the scalar metric perturbations, it turns out that one can express 
the resulting action in terms of a variable $v$ which is the 
coordinate-invariant scalar matter field fluctuation \cite{Mukh85}
\begin{equation}
v \, = \, a \bigl( \delta \varphi + 
{({\varphi^{(0)})^{ \prime}} \over {\cal{H}}} \Phi \bigr) \, .
\end{equation}
Here, $\varphi^{(0)}$ is the background scalar matter field, $\delta \varphi$
is the scalar field fluctuation, a prime denotes the derivative with 
respect to conformal time
$\eta$ ($\eta$ being defined by $dt = ad\eta$), 
${\cal H} \, = \, a^{\prime} / a$, and $\Phi$ is the generalized Newtonian
gravitational potential (in a coordinate system in which the metric
tensor - including fluctuations - is diagonal, $2 \Phi$ is the
perturbation of the time-time component of the metric). It
can then be shown that the action $S_2$ for the fluctuations reduces 
to the action of a single gauge invariant free scalar field (namely $v$) 
with a time dependent mass \cite{Mukh88,Sasaki}  (the time dependence 
reflects the expansion of the background space-time) 
\begin{equation} \label{quad}
S_2 \, = \, {1 \over 2} \int dt d^3x \bigl( v^{\prime 2} - (\nabla v)^2 + {{z^{\prime \prime}} \over z} v^2 \bigr) \, ,
\end{equation}
where the function
\begin{equation}
z \, = \, {{a \varphi_0^{\prime}} \over {\cal H}} \, ,
\end{equation}
a function of the background fields, determines the effective mass for the
field $v$.
The action $S_2$ has the same form as the action for a free scalar matter 
field in a time dependent gravitational or electromagnetic background, and 
we can use standard methods to quantize this theory (see e.g. \cite{Birrell}).
Each momentum mode of the field obeys a harmonic oscillator equation
with time dependent mass. The time dependence of the mass is reflected in 
the nontrivial form of the 
solutions of the mode equations. Let us now follow the evolution of
a particular mode from the beginning of inflation until late times.
The mode starts out in its vacuum state when its wavelength is smaller
than the Hubble radius. While the wavelength remains smaller than the
Hubble radius, the mode functions oscillate since the time dependence
of the mass is negligible, as can be seen from (\ref{quad}). However, once
the mode crosses the Hubble radius, the spatial gradient term becomes
negligible and the mass term begins to dominate. The mode function thus
no longer oscillates, but its amplitude begins to grow as $v \sim z$.
This corresponds to squeezing of the vacuum state and corresponds to
the generation of fluctuations. This calculation can be followed
up to the time the mode re-enters the Hubble radius at late times.
It turns out (see Section 5) that the amplitude of the resulting density
generically is predicted to be several orders of magnitude larger than
what is compatible with current observations, unless some parameter in
the quantum field theory Lagrangian is set to a value much smaller than
would be inferred from dimensional analysis.

\section{Problems of Inflationary Cosmology}

In spite of the spectacular success of the inflationary scenario in
predicting a spectrum of microwave anisotropies and large-scale
density fluctuations in excellent agreement with the recent observation,
scalar field-driven inflationary models suffer from some serious
conceptual problems.

The main success of inflationary cosmology is that it provides a causal
theory for the generation of large-scale cosmological fluctuations. However,
this success directly leads to a major problem
for most realizations of scalar field-based models of inflation studied 
up to now. It concerns the amplitude of the density perturbations which are 
induced by quantum fluctuations during the period of accelerated expansion 
as discussed in the previous section. Unless a parameter in the
scalar field potential is set to have a value several orders of magnitude
smaller than what would be given by dimensional analysis, the models
predict an amplitude of the fluctuation spectrum several order of
magnitude larger than the predicted amplitude. For example, in a
model with a single inflaton field with quartic potential, the
quartic coupling constant $\lambda$ must be of the order of $10^{-12}$
in order that the resulting amplitude of fluctuations agrees with 
observations. This situation is clearly unsatisfactory for a cosmological
scenario motivated by the desire to eliminate cosmological fine-tunings.
There have been many attempts to justify such small parameters based on 
specific particle physics models, but no single convincing model has emerged.
However, this is probably the least serious of the problems mentioned here.

In many models of inflation, in particular in chaotic inflation, the period 
of inflation is so long that comoving scales of cosmological interest today 
correspond to a physical wavelength much smaller than the Planck length at 
the beginning of inflation. In extrapolating the evolution of cosmological 
perturbations according to linear theory to these very early times, one is 
implicitly making the assumption that the theory remains perturbative to 
arbitrarily high energies, and that the classical theory of general
relativity remains the appropriate framework for describing space-time.
Both of these assumptions are clearly not justified. It has recently been
shown that some (admittedly quite violent) changes to the physics on
length scales smaller than the Planck length can lead to a spectrum
of density fluctuations totally different from what the usual theory
predicts \cite{Martin}. This can be called the ``trans-Planckian problem''
for inflationary cosmology. On the other hand it shows that in an 
inflationary Universe, the spectrum of fluctuations can potentially be
used to explore Planck-scale physics, thus turning the ``problem'' into
a ``window'' of opportunity. Planck-scale physics may 
not only alter the spectrum of fluctuations, it can also dramatically alter 
the background cosmology, as is seen in ``Pre-big-bang Cosmology" \cite{PBB},
a string-motivated dilaton-gravity model  which undergoes a
dilaton-dominated phase of super-exponential expansion.

Scalar field-driven inflation does not eliminate singularities from 
cosmology. Although the standard assumptions of the Penrose-Hawking 
theorems (the theorems which is the context of Einstein gravity coupled
to classical fluid matter show that an initial cosmological singularity
is inevitable) break down if matter has an equation of state with negative 
pressure, as is the case during inflation, nevertheless it can be shown 
that an initial singularity persists in inflationary cosmology \cite{Borde}. 
This implies that the theory is incomplete. In particular, the physical 
initial value problem is not defined.
 
The Achilles heel of our current inflationary models in without doubt the
``cosmological constant problem''. There is some as of yet unknown
mechanism which prevents the bare cosmological constant, which
in theories with quantum fields is predicted to be at least 62 orders
of magnitude larger than the observational limit (this number comes from
assuming the cancellation of vacuum energies on scales larger than the
supersymmetry breaking scale taken to be about $1$TeV), from gravitating.
How do we know that this unknown mechanism does not also lead the
transient ``cosmological constant'' given by the potential energy
of the scalar field to be gravitationally inert, thus eliminating the
basis of scalar field-driven inflation? 

\section{Future Directions}

In the light of the problems of scalar field potential-driven inflation 
discussed in 
the previous sections, many cosmologists have begun thinking about new 
avenues towards early Universe cosmology which, while maintaining (some of) 
the successes of inflation, address and resolve some of its difficulties. 
In the same way that inflationary cosmology builds on and transcends
standard big bang cosmology by making use of a new theory of matter
(quantum field theory in the case of inflation), it is
quite likely that a resolution of the problems of inflationary cosmology
will once again come from an improved description of matter at high
energies. The main candidate at the moment for such a theory is
string theory. String theory, in fact, will also lead to a modified
picture of space-time at short distances, and may  allow a unified
quantum description of both the background space-time and of the
cosmological fluctuations. 

There are at least two ways in which an improved cosmological model could
come about. It is possible that a better understanding of string theory
will lead to a convincing realization of inflation which does not suffer
from the problems mentioned in the previous section, and that many
of the cosmological predictions of inflation for present day observations
would remain unchanged. However, it is also possible that string theory
will provide alternatives to inflationary cosmology, a possibility
which would lead to clear observational signatures.

There are several reasons to hope that string theory might lead to
an improved realization of inflation. One reason is that the vacuum
space of string theory is a complicated moduli space, and the individual
directions in this space (e.g. radii of extra dimensions) can be
associated with scalar fields with potentials which vanish at the
perturbative level. It is hoped that nonperturbative corrections might
generate in a natural way the flat potentials required for inflation.
The realization that string theory contains higher dimensional fundamental
objects called branes has led to speculations that our space-time
might be a brane in a higher dimensional space-time. These
``brane-world'' scenarios have opened new avenues to realizing inflation
in the context of string theory (see e.g. \cite{Burgess} for a 
recent attempt and for references to earlier work).

However, it is also possible that string theory will generate
an alternative to inflationary cosmology which maintains most of
the successes of inflation, but at the same time gives rise to
predictions with which this new theory can be distinguished from
inflation. One approach which 
has received a lot of recent attention is pre-big-bang cosmology \cite{PBB}, 
a theory in which the Universe starts in an empty and flat dilaton-dominated 
phase which leads to super-exponential expansion. A nice feature of this 
theory is that the mechanism of acceleration is completely independent of 
a scalar field potential 
and thus independent of the cosmological constant issue. Pre-big-bang
cosmology in its simplest realizations does not give rise to a spectrum
of scale-invariant adiabatic fluctuations. Rather, the spectrum is
isocurvature (see e.g. \cite{GV} for a recent review and original
references). Another recent attempt to provide an alternative to
inflation in the context of string theory is the ``Ekpyrotic Universe''
scenario \cite{KOST}, a nonsingular cosmological model designed to solve 
the problems of SBB cosmology mentioned in Section 1 and to provide
a spectrum of scale-invariant adiabatic perturbations without any period of 
acceleration (see, however, \cite{KKL,Lyth,BF} for criticisms
of this scenario).

It is, however, also possible that some of the problems of inflationary
cosmology mentioned in Section 5 can be addressed within the context
of more conventional physics (general relativity plus quantum field
theory). In particular, it is possible that some of the problems
are consequences of neglecting the intrinsically nonlinear structure
of general relativity (see e.g. \cite{RHBcosmo} for some speculations
along these lines). 

\acknowledgments{I am grateful to Douglas Scott for
providing Figures 1 and 7. This work was supported in part by the 
US Department of Energy under Contract DE-FG02-91ER40688, Task A.}

%\bibliography{biblio}

\begin{thebibliography}{99}

\bibitem{Scott} G. Smoot and D. Scott, in ``Reviews of Particle Properties'',
in press (2001).
 
\bibitem{Gush} H. Gush, M. Halpern and E. Wishnow, Phys. Rev. Lett. {\bf 65},
537 (1990).

\bibitem{COBE} J. Mather et al., Astrophys. J. {\bf 420}, 439 (1994).

\bibitem{RHB}
%\cite{Brandenberger:1999sw}
R.~H.~Brandenberger,
``Inflationary cosmology: Progress and problems,''
arXiv:hep-ph/9910410.
%%CITATION = HEP-PH 9910410;%%

%\cite{Guth:1981zm}
\bibitem{Guth}
A.~H.~Guth,
%``The Inflationary Universe: A Possible Solution To The Horizon And Flatness Problems,''
Phys.\ Rev.\ D {\bf 23}, 347 (1981).
%%CITATION = PHRVA,D23,347;%%

\bibitem{Mukh80}
%\cite{Mukhanov:1981xt}
V.~F.~Mukhanov and G.~V.~Chibisov,
%``Quantum Fluctuation And 'Nonsingular' Universe. (In Russian),''
JETP Lett.\  {\bf 33} (1981) 532
[Pisma Zh.\ Eksp.\ Teor.\ Fiz.\  {\bf 33} (1981) 549].
%%CITATION = JTPLA,33,532;%%

\bibitem{Lukash} V. Lukash, Pis'ma Zh. Eksp. Teor. Fiz. {\bf 31}, 631 (1980).

\bibitem{Linde}
%\cite{Linde:1983gd}
A.~D.~Linde,
%``Chaotic Inflation,''
Phys.\ Lett.\ B {\bf 129}, 177 (1983).
%%CITATION = PHLTA,B129,177;%%

\bibitem{Hu} W. Hu, http://background.uchicago.edu/ .

\bibitem{MFB}
%\cite{Mukhanov:1992me}
V.~F.~Mukhanov, H.~A.~Feldman and R.~H.~Brandenberger,
%``Theory of cosmological perturbations. Part 1. Classical perturbations. Part 2. Quantum theory of perturbations. Part 3. Extensions,''
Phys.\ Rept.\  {\bf 215}, 203 (1992).
%%CITATION = PRPLC,215,203;%%


\bibitem{Mukh85}
%\cite{Mukhanov:1985rz}
V.~F.~Mukhanov,
%``Gravitational Instability Of The Universe Filled With A Scalar Field,''
JETP Lett.\  {\bf 41}, 493 (1985)
[Pisma Zh.\ Eksp.\ Teor.\ Fiz.\  {\bf 41}, 402 (1985)].
%%CITATION = JTPLA,41,493;%%

\bibitem{Mukh88}
%\cite{Mukhanov:1988jd}
V.~F.~Mukhanov,
%``Quantum Theory Of Gauge Invariant Cosmological Perturbations,''
Sov.\ Phys.\ JETP {\bf 67}, 1297 (1988)
[Zh.\ Eksp.\ Teor.\ Fiz.\  {\bf 94N7}, 1 (1988)].
%%CITATION = SPHJA,67,1297;%%

\bibitem{Sasaki}
%\cite{Sasaki:1986hm}
M.~Sasaki,
%``Large Scale Quantum Fluctuations In The Inflationary Universe,''
Prog.\ Theor.\ Phys.\  {\bf 76}, 1036 (1986).
%%CITATION = PTPKA,76,1036;%%

\bibitem{Birrell} N. Birrell and P.C.W. Davies, ``Quantum fields
in curved space'' (Cambridge Univ. Press, Cambridge, 1982).

\bibitem{Martin}
%\cite{Martin:2001xs}
J.~Martin and R.~H.~Brandenberger,
%``The trans-Planckian problem of inflationary cosmology,''
Phys.\ Rev.\ D {\bf 63}, 123501 (2001)
[arXiv:hep-th/0005209];\\
%%CITATION = HEP-TH 0005209;%%
%\cite{Brandenberger:2001wr}
R.~H.~Brandenberger and J.~Martin,
%``The robustness of inflation to changes in super-Planck-scale physics,''
Mod.\ Phys.\ Lett.\ A {\bf 16}, 999 (2001)
[arXiv:astro-ph/0005432].
%%CITATION = ASTRO-PH 0005432;%%

\bibitem{PBB}
%\cite{Gasperini:1993em}
M.~Gasperini and G.~Veneziano,
%``Pre - big bang in string cosmology,''
Astropart.\ Phys.\  {\bf 1}, 317 (1993)
[arXiv:hep-th/9211021].
%%CITATION = HEP-TH 9211021;%%

\bibitem{Borde}
%\cite{Borde:1994xh}
A.~Borde and A.~Vilenkin,
%``Eternal inflation and the initial singularity,''
Phys.\ Rev.\ Lett.\  {\bf 72}, 3305 (1994)
[arXiv:gr-qc/9312022].
%%CITATION = GR-QC 9312022;%%

\bibitem{Burgess}
%\cite{Burgess:2001fx}
C.~P.~Burgess, M.~Majumdar, D.~Nolte, F.~Quevedo, G.~Rajesh and R.~J.~Zhang,
``The inflationary brane-antibrane universe,''
JHEP {\bf 0107}, 047 (2001)
[arXiv:hep-th/0105204].
%%CITATION = HEP-TH 0105204;%%

\bibitem{GV}
%\cite{Veneziano:2000pz}
G.~Veneziano,
``String cosmology: The pre-big bang scenario,''
arXiv:hep-th/0002094.
%%CITATION = HEP-TH 0002094;%%

\bibitem{KOST}
%\cite{Khoury:2001wf}
J.~Khoury, B.~A.~Ovrut, P.~J.~Steinhardt and N.~Turok,
``The ekpyrotic universe: Colliding branes and the origin of the hot big  bang,''
arXiv:hep-th/0103239.
%%CITATION = HEP-TH 0103239;%%

\bibitem{KKL}
%\cite{Kallosh:2001ai}
R.~Kallosh, L.~Kofman and A.~D.~Linde,
``Pyrotechnic universe,''
arXiv:hep-th/0104073.
%%CITATION = HEP-TH 0104073;%%

\bibitem{Lyth}
%\cite{Lyth:2001pf}
D.~H.~Lyth,
``The primordial curvature perturbation in the ekpyrotic universe,''
arXiv:hep-ph/0106153.
%%CITATION = HEP-PH 0106153;%%

\bibitem{BF}
%\cite{Brandenberger:2001bs}
R.~Brandenberger and F.~Finelli,
``On the spectrum of fluctuations in an effective field theory of the  ekpyrotic universe,''
arXiv:hep-th/0109004.
%%CITATION = HEP-TH 0109004;%%

\bibitem{RHBcosmo}
%\cite{Brandenberger:1999su}
R.~H.~Brandenberger,
``Back reaction of cosmological perturbations,''
arXiv:hep-th/0004016.
%%CITATION = HEP-TH 0004016;%%

\end{thebibliography}

\end{document}